\documentclass[12pt]{article}
\usepackage[sumlimits,intlimits,namelimits]{amsmath}
\usepackage{amssymb}
\usepackage[english]{babel}
\usepackage{bbm}
\usepackage{parskip}
\global\parskip 6pt
\newcommand{\be}{\begin{eqnarray}}
\newcommand{\ee}{\end{eqnarray}}

\begin{document}
\begin{titlepage}
\begin{flushright}
LMU-ASC 06/16\\
\vspace{1.5cm}
October 8, 2015\\
\end{flushright}
\vspace{20mm}

\begin{center}
{\Large \bf 
Open Superstring Field Theory on the Restricted Hilbert Space}
\\ 
\end{center}

\vspace{25mm}
\begin{center}

{\bf
Sebastian Konopka\footnote{\texttt{sebastian.konopka@physik.uni-muenchen.de}}  
and
Ivo Sachs\footnote{\texttt{ivo.sachs@physik.uni-muenchen.de}}}

\vspace{5mm}

\footnotesize{
{\it Arnold Sommerfeld Center for Theoretical Physics, Ludwig-Maximilians Universit\"at\\
Theresienstra\ss{}e 37, D-80333 M\"unchen, Germany}}

\end{center}
\vspace{25mm}
\begin{abstract}
\baselineskip=14pt
Recently an action for open superstring field theory was proposed where the Neveu-Schwarz sector is formulated in the large Hilbert space while the Ramond sector lives in a restriction of the small Hilbert space. The purpose of this note is to clarify the relation of the restricted Hilbert space with other approaches and to formulate open superstring field theory entirely in the small Hilbert space. 
\end{abstract}

\end{titlepage}
\section{Introduction}
The problem of formulating an action for interacting covariant open superstring field theory has a long history, starting with Witten's cubic action \cite{Witten:1986qs} which used an unconstrained string field. %containing all powers of the superconformal ghost zero modes. 
This cubic theory has two short comings: One problem is the presence of singularities in the Neveu-Schwarz (NS) sector due to collisions of picture changing operators.  Another issue is that the kinetic term (more precisely the inner product) is degenerate in the Ramond (R) sector. The first problem can be remedied by smearing out the picture changing operator \cite{Erler:2013xta} (see also \cite{Iimori:2013kha} for earlier work in this direction). This results in a consistent (although non-polynomial) BV-action for the NS sector of open superstring field theory on the small Hilbert space. On the other hand, an action for the NS sector in the large Hilbert space has been formulated long time ago by Berkovits \cite{Berkovits:1995ab}. This theory is attractive due its simple form and is well suited for explicit calculations (e.g. \cite{Erler:2013wda}) but its BV-quantization is less clear. However, recently it has been shown that Berkovits' theory is related to the BV-action on the small Hilbert space by a field redefinition \cite{Erler:2015rra,Erler:2015uba}. This shows that the former does indeed realize a decomposition of the supermoduli space. Furthermore, it was shown in \cite{Konopka:2015tta} that the non-polynomial BV-action \cite{Erler:2013xta} (and thus the Berkovits action) does reproduce the perturbative tree-level S-matrix to all orders. 

For the combined theory of NS- and R- sectors consistent (i.e. gauge invariant) field equations have been formulated in \cite{Erler:2015lya} and shown to produce the correct tree-level S-matrix elements \cite{Konopka:2015tta}  but, due to the lack of cyclicity, of the multi-string vertices these field equations cannot derive from an action. Furthermore, the above-mentioned issue with the kinetic term in the Ramond (R) sector was not addressed in \cite{Erler:2015lya}. On the other hand, in  \cite{Jurco:2013qra} and  \cite{Kunitomo:2015usa} the degeneracy of the Ramond kinetic term was avoided with the help of a suitable restriction of the Ramond Hilbert space. Indeed, it was noticed \cite{Sazdovic:1987yz} in the early days of string field theory that Witten's theory propagates only a subset of constrained string fields \cite{Kazama:1986cy}-\cite{Date:1986tg}. This was subsequently related to the presence of an extra gauge symmetry (not generated by the BRST charge) that can be fixed to remove all fields that do not satisfy the constraint \cite{Kugo:1988mf} (see also \cite{Kohriki:2012pp}).

A gauge invariant action for the interacting theory was recently proposed in \cite{Kunitomo:2015usa} (see also \cite{Matsunaga:2015kra}) with smeared picture changing operators and Ramond fields in the restricted Hilbert space. The above problem with cyclicity of the vertices was avoided by taking the the NS field to live in the large Hilbert space akin to the Berkovits formulation. On the other hand, in \cite{Jurco:2013qra} a geometric approach, based on the decomposition of the supermoduli space was outlined, which is formulated in the small Hilbert space with a constrained Ramond sector. Furthermore, in \cite{Sen:2015uaa} another geometric construction was proposed where the restriction on the Ramond fields is substituted by the introduction of auxiliary fields \footnote{In fact, the proposals  \cite{Jurco:2013qra} and \cite{Sen:2015uaa} were worked out for the closed type II superstring but the idea is easily adapted to the open string.}.  

The purpose of this note is twofold. First we clarify the relation between the restricted and unrestricted Ramond Hilbert spaces. In particular, we show explicitly that the restrictions used in  \cite{Kunitomo:2015usa} and \cite{Jurco:2013qra} are the same and furthermore that the cohomology of the restricted Hilbert space is the same as that of the unrestricted space. The latter result was previously obtained in \cite{Henneaux:1987ux}\footnote{We would like to thank Y. Okawa for pointing out this reference to us.}. In the second part we propose a modification of the construction \cite{Erler:2015lya} for the R-NS vertices which is cyclic in the small, restricted Hilbert space. Our construction applies straight forwardly to the open string version of the formalism working with auxiliary fields in \cite{Sen:2015uaa} thus providing an algebraic construction of the corresponding vertices for open strings. Then, invoking the result of \cite{Konopka:2015tta} one concludes that this construction reproduces the correct perturbative tree-level S-matrix. On the other hand, 
our construction provides a classical action for the open superstring in the small, restricted Hilbert space, provided the picture changing operators used in \cite{Jurco:2013qra, Kunitomo:2015usa} can be defined in a way that is compatible with the interaction vertices.

% on the small Hilbert space our construction gives rise to a classical action for the open superstring which is entirely formulated in the small Hilbert space. %In particular, this completes the proposal outlined in \cite{Jurco:2013qra} for the case of the open superstring. 
%On the other hand, our construction applies straight forwardly to the open string version of the formalism working with auxiliary fields in \cite{Sen:2015uaa} thus providing an algebraic construction of the corresponding vertices for open strings. Then, invoking the result of \cite{Konopka:2015tta} one concludes that these constructions reproduce the correct perturbative S-matrix. 
%
\section{Restricted Hilbert Space}
Let us start with the restricted Ramond Hilbert space spanned by vectors of the form \cite{Kunitomo:2015usa}- \cite{Kohriki:2012pp}
\begin{equation}\label{rs1}
\psi = \phi_1|\!\downarrow\rangle+\,\gamma_0 \phi_2|\!\downarrow\rangle-(-1)^{|\phi_1|}G_0 \phi_2|\!\uparrow\rangle
\end{equation}
where $|\!\downarrow\rangle=b_0|\!\uparrow\rangle$, $|\phi|$ denotes the Grassman parity of $\phi$, $\gamma_0$ is the zero mode of the commuting superconformal ghost and $G_0$ the (matter plus ghost) supercharge with the $\gamma_0b_0$ contribution subtracted. More concretely, we decompose the BRST charge $Q$ as 
\begin{equation}
Q=c_0L_0+b_0M+\gamma_0G_0 +\beta_0K-\gamma_0^2b_0+\tilde Q
\end{equation}
where $L_0,M,G_0,K,\tilde Q$ have no dependence on the ghost zero modes (see e.g. \cite{Kohriki:2012pp} for details). Then, using that $\{\tilde Q,G_0\}=0$ and $G_0^2=L_0$ it is not hard to see that 
\begin{eqnarray}
Q\psi&=&\left(M(G_0\phi_2)+K(\phi_2) +\tilde Q(\phi_1) \right)|\!\downarrow\rangle+ \,\gamma_0\left(G_0(\phi_1)+\tilde Q(\phi_2)\right)|\!\downarrow\rangle \nonumber\\
&&+\,(-1)^{|\phi_1|}G_0\left(G_0(\phi_1)+ \tilde Q(\phi_2) \right)|\!\uparrow\rangle\,.
\end{eqnarray}
According to \cite{Kohriki:2012pp}, $\phi_2$ can be gauged away completely\footnote{Notice however, that there are some subtleties when $G_0 \phi_2 = 0$.}. The closedness condition reduces then to 
\begin{equation}
\tilde Q \phi_1=G_0\phi_1=0\,,
\end{equation}
with a residual gauge freedom
\begin{equation}
\delta_\lambda\phi_1=\tilde Q\lambda\;,\quad G_0\lambda=0\,.
\end{equation}

Let us now compare this with the cohomology of the unrestricted Ramond sector. Because the cohomology of $Q$ is known to be isomorphic to the relative cohomology $H_{rel}^\bullet(Q)$ calculated on on the subspace defined by $b_0\psi=\beta_0\psi=0$ \cite{Henneaux:1987ux,FigueroaO'Farrill:1988hu} we consider this case. A generic vector in this subspace is given by $\psi=\phi\,|\!\downarrow\rangle$ with $\phi$ independent of $\gamma_0$ and $c_0$.  Then, $Q\psi=0$ reduces to 
 \begin{equation}
\tilde Q \phi=G_0\phi=0\,,
\end{equation}
with the same residual gauge freedom as above. Thus the cohomology of the restricted Ramond sector (\ref{rs1}) agrees with that of the unrestricted Ramond Hilbert space as previously shown in \cite{Henneaux:1987ux}. 

Next, we compare the restriction (\ref{rs1}) with the approach of \cite{Jurco:2013qra}. The constraint, originally formulated in \cite{Yeh:1993qc}, is again motivated by imposing that the cokernel of the picture changing operator\footnote{Note that there is no well-established algebraic characterization of the $(-\frac{1}{2})$ states in terms of the modes of $\beta$ and $\gamma$. For (\ref{x0}), one possible choice is to require that $\beta_k^{n_k} | \psi \rangle = \gamma_l^{m_l} | \psi \rangle = 0$ for $l > 0$ and $k \geq 0$ and natural numbers $n_k$ and $m_l$. This is not a problem for free string field theory but becomes an issue in the presence of interaction vertices which generically do not preserve this definition.}

\begin{equation}\label{x0}
X_0=(G_0-2\gamma_0b_0)\delta(\beta_0) +b_0\delta'(\beta_0)
\end{equation}
vanishes. This leads to the condition 
\begin{equation}\label{ryeh1}
\beta_0^2\psi=0
\end{equation}
with general solution, 
\begin{equation}
\psi=\phi_1^{(0)}|\!\downarrow\rangle+ \gamma_0\phi_1^{(1)}|\!\downarrow\rangle+\phi_2^{(0)}|\!\uparrow\rangle+\gamma_0\phi_2^{(1)}|\!\uparrow\rangle
\end{equation}
where $\phi_i^{(j)}$ are independent of $\gamma_0$ and $c_0$. Now requiring that the condition (\ref{ryeh1}) is preserved by $Q$ implies that $\phi_2^{(1)}=0$ and $\phi_2^{(0)}=-(-1)^{|\phi_1^{(1)}|}G_0\phi_1^{(1)}$ and thus (\ref{ryeh1}) and (\ref{rs1}) define the same invariant subspace. 
Finally we note that every vector in this subspace can be written as $\psi = X_0\tilde\psi$, where $\tilde \psi$ is an arbitrary string field with picture $-\frac{3}{2}$.  This follows from the identities   \cite{Yeh:1993qc}
\begin{eqnarray}
\delta(\gamma_0) &=& | 0, - \frac{3}{2} \rangle \langle 0, -\frac{3}{2} | \\
\delta(\beta_0)&=&|0,-\frac{1}{2}\rangle\langle 0,-\frac{1}{2}|,\\
\delta'(\beta_0)&=&-|0,-\frac{1}{2}\rangle\langle 1,-\frac{1}{2}|+|1,-\frac{1}{2}\rangle\langle 0,-\frac{1}{2}|
\end{eqnarray}
where the index $-\frac{1}{2}$ resp. $-\frac{3}{2}$ denotes the picture and $|n,-\frac{1}{2}\rangle=\gamma_0^n|0,-\frac{1}{2}\rangle$. Then, for $\tilde\psi=\phi_1|\!\downarrow\rangle+\phi_2|\!\uparrow\rangle$ with $\phi_i=\sum\limits_{n=0}^\infty\beta_0^n\phi_i^{(n)}\delta(\gamma_0)$ we find
\begin{eqnarray}
X_0\tilde\psi=\left(G_0(\phi_1^{(0)})-(-1)^{|\phi_2|}\phi_2^{(1)}\right)|\!\downarrow\rangle-(-1)^{|\phi_2|}\gamma_0\phi_2^{(0)}|\!\downarrow\rangle+G_0\phi_2^{(0)}|\!\uparrow\rangle
\end{eqnarray}
where we have used that $\delta(\gamma_0)\delta(\beta_0) = |0, -\frac{3}{2} \rangle\langle 0, -\frac{1}{2}|$. We then see that $X_0\tilde\psi$  is indeed of the form (\ref{rs1}) with 
\begin{eqnarray}
\phi_1=G_0(\phi_1^{(0)})-(-1)^{|\phi_2|}\phi_2^{(1)}\qquad\hbox{and}\qquad\phi_2=(-1)^{|\phi_2|}\phi_2^{(0)}\,.
\end{eqnarray}

\section{Open Superstring Field Theory in the restricted Hilbert space}
In \cite{Erler:2013xta,Erler:2015lya} the NS sector of open superstring theory was obtained as a gauge transformation of the free theory, where the gauge transformation is defined  as a hierarchy of gauge products on the large Hilbert space with each gauge product obtained from lower order products by means of a homotopy for the nilpotent operator $\eta_0$, %that is the zero mode of the fermion 
that enters in the bosonization of the superconformal ghost $\gamma(z)$. In  \cite{Erler:2013xta} a class of homotopies built out of 
\begin{equation}
 \xi=\oint\frac{dz}{2\pi i} f(z) \xi(z)
\end{equation}
was considered, where $f(z)$ is required to be holomorphic in some annulus that contains the unit circle and $\xi(z)$ enters in %is the fermion entering 
the bosonization of superconfomal ghost, $\beta(z)=\partial\xi(z)e^{-\phi(z)}$. 
%However, the construction works equally well for a larger class of homotopies. In particular, one can take  $\xi=\Theta(\beta_0)$ which is the natural choice in view of the restricted Hilbert space (\ref{rs1}).  
Furthermore, the homotopy was taken to be the same irrespective of whether the string products defining the string vertices through 
\begin{equation}
C_n(\Psi_1,\cdots,\Psi_n)= \omega(\Psi_1,M_{n-1}(\Psi_2,\cdots,\Psi_{n-1}))
\end{equation}
has zero or one Ramond input \cite{Erler:2015lya}. Here, $\Psi$ denotes a combined string field in the R- and NS-sector. However,  this is not required by gauge-invariance. To illustrate this we consider the string product
\begin{eqnarray}\label{M2cyc}
M_2=\frac{1}{3}\{{} X,m_2\}P_2^{<0>}+ Xm_2 P_2^{<1>}+m_2 P_2^{<2>}
\end{eqnarray}
where $P_2^{<n>}$ is the projector on $n$ Ramond inputs among the two inputs of $m_2$ and   $m_2=*$ is Witten's string product. The picture changing operator, $X$ is related to $\xi$ through the graded commutator, $X=[Q,\xi]$. Finally,  $\{{} X,m_2\}$ is the graded anti-commutator of $ X$ and $m_2$. For zero Ramond inputs $M_2$ is cyclic with respect to the standard symplectic form by construction since the combination $\{X,m_2\}$ sums over all possible insertions of a picture changing operator (see \cite{Erler:2013xta} for details and notation).  %In particular, $M^{<1>}=Xm_2P_2^{<2>}$.
%Then, since $XYX=Y$, where $Y=c_0\delta'(\gamma_0)$ is the inverse picture changing operator, the identity  $XY \psi=\psi$ holds in the restricted Ramond sector (\ref{rs1}) (see also \cite{Kohriki:2012pp}). Then, with the assignment of picture changing operators as in (\ref{M2cyc}) the three vertex is now cyclic for Ramond insertions as well. Indeed, 
For vertices involving two Ramond fields we have 
\begin{eqnarray}\label{c2}
\omega(N,M_2(R,R))&=&\omega(N,m_2(R,R))=\omega(R,m_2(R,N))%\\
%&=&\omega(XY R,m_2(R,R))=\omega(YR,Xm_2(R,N))=\omega(YR,XYXm_2(R,N))\nonumber\\
%&=&\omega(XYR,YXm_2(R,N))=\omega(R,YXm_2(R,N))=\omega(R,YM_2(R,N))\nonumber
\end{eqnarray}
where %$\omega(\cdot,Y\cdot)$ is the non-degenerate symplectic form in the restricted Ramond sector and 
$N$ and $R$ denote NS- and R- string fields respectively. At first sight it looks as if $M_2$ were not cyclic since there is an $X$ missing in front of $m_2$ on the right hand side of (\ref{c2}). However, we will see in the end that this is exactly what we need, because of subtleties in defining a symplectic form on the R-string fields.

Next, let us consider the 4-vertex. First, we have from (\ref{M2cyc}) % [[[ define the homotopy]]]]  (\ref{M2cyc})
\begin{equation}
[M_2,M_2](R,R,R)=2Xm_2\circ m_2(R,R,R) =0
\end{equation}
due to associativity of the star product ($m_2\circ m_2=0$). Thus, to this order the $A_\infty$ consistency condition (or equivalently the BV-equation) allows us to set $M_3(R,R,R)=0$. For two Ramond inputs we have 
\begin{eqnarray}
\frac{1}{2}[M_2,M_2](R,N,R)&=&m_2\circ Xm_2(R,N,R)=-[Q,[m_2,\mu_2]](R,N,R)\,,\nonumber%=: -[Q,M_3](R,N,R)
\end{eqnarray}
where
\begin{equation}\label{mu2}
\mu_2=\xi m_2 P_2^{<1>}+\frac{1}{3}\{{} \xi,m_2\}P_2^{<0>}\,.
\end{equation}
Since the gauge products $\mu_n$ never have more than one Ramond input \cite{Erler:2015lya}, the $A_\infty$ consistency condition, $\frac{1}{2}[M_2,M_2]+[Q,M_3]=0$, then fixes $M_3$ completely as
\begin{equation}
M_3(R,N,R)=m_3(R,N,R)\,,
\end{equation}
where $m_3=[m_2,\mu_2]$ and we have used associativity of $m_2$.  Associativity then also implies  
that $\eta M_3(R,N,R)=-\eta [m_2,\mu_2](R,N,R) =0$ and thus  $M_3$ is in the small Hilbert space. 

Similarly, for one Ramond input
\begin{eqnarray}
\frac{1}{2}[M_2,M_2](N,R,N)&=&Xm_2\circ Xm_2(N,R,N)=-\frac{1}{2}[Q,[Xm_2P_2^{<1>},\mu_2P_2^{<1>}]](N,R,N)\nonumber\\
&=&-\frac{1}{2}[Q,[M_2,\mu_2P_2^{<1>}](N,R,N)=-\frac{1}{2}[Q,[M_2,\mu_2]](N,R,N)\,.%\\
%&=:& -[Q,M_3](N,R,N)\nonumber
\end{eqnarray}
To continue we choose the homoptopy for $\eta$ defining the gauge product $\mu_3$ as 
\begin{eqnarray}
\mu_3=\frac{1}{4}\{{} \xi,m_3\}P_3^{<0>}+\xi m_3 P_3^{<1>}\,.
\end{eqnarray}
Then,
\begin{eqnarray}
\mu_3(N,R,N)=\xi m_3(N,R,N)=\xi m_2\circ \xi m_2(N,R,N)\,.
\end{eqnarray}
Using, associativity of $m_2$ again we then find 
\begin{eqnarray}
M_3(N,R,N)&=&\frac{1}{2}\left([M_2,\mu_2]+[Q,\mu_3]\right)(N,R,N)\nonumber\\
&=&M_2^{<1>}\mu_2(N,R,N)=X m_2^{<1>}\mu_2(N,R,N)=Xm_3P_3^{<1>}(N,R,N)
\end{eqnarray}
which is in the small Hilbert space. More generally, for a generic permutation of the R- and N inputs 
\begin{equation}\label{3gen}
M_3P_3^{<1>}=X m_2^{<1>}\mu_2P_3^{<1>}= Xm_3P_3^{<1>}\,
\end{equation}
holds. Thus, modulo the factor $X$ that will be dealt with below, proving cyclicity of $M_3$ is reduced to show cyclicity of $m_3$. 
%The extra factor of $X$ in (\ref{3gen}) then arises from the insertion of $Y$ in the symplectic form in the Ramond sector as in (\ref{c2}). 
Explicitly, we have 
\begin{eqnarray}\label{cyc1}
\omega(N_1,M_3(R_1,N_2,R_2))&=&\omega(N_1,m_3(R_1,N_2,R_2))\\
&=&\omega_L(N_1,\xi_0m_2(\xi m_2(R_1,N_2),R_2))+\omega_L(N_1,\xi_0 m_2(R_1,\xi m_2(N_2,R_2)))\,,\nonumber
\end{eqnarray}
where $\omega_L$ is the symplectic form evaluated in the large Hilbert space and which reproduces the symplectic form, $\omega$, on the small Hilbert space upon insertion of the zero mode $\xi_0$ \cite{Erler:2013xta}.  Now, commuting $\xi_0$ through to $R_1$ and using cyclicity of $m_2$ we get 
\begin{eqnarray}\label{3cyc1}
\omega(N_1,M_3(R_1,N_2,R_2))&=&\omega_L(\xi m_2(\xi_0R_1,N_2), m_2(R_2,N_1)) +\omega_L(\xi_0R_1, m_2(\xi m_2(N_2,R_2),N_1))\,.\nonumber
\end{eqnarray}
Since $\xi$ is BPZ-even we  then have
\begin{eqnarray}\label{cyc3}
\omega(N_1,M_3(R_1,N_2,R_2))&=&\omega_L( m_2(\xi_0R_1,N_2),\xi m_2(R_2,N_1))+\omega_L(\xi_0R_1, m_2(\xi m_2(N_2,R_2),N_1))  \nonumber\\
&=&\omega_L( \xi_0R_1, m_2(N_2, \xi m_2(R_2,N_1)))+\omega_L(\xi_0R_1, m_2(\xi m_2(N_2,R_2),N_1))\nonumber\\
%&=&\omega_L(\xi_0R_1, m_2(N_2, \xi m_2(R_2,N_1)))+\omega_L(\xi_0R_1, m_2(\xi m_2(N_2,R_2),N_1))\nonumber\\
&=&\omega(R_1,m_3(N_2,R_2,N_1))\,.\nonumber\\
%&=&\omega(XY R_1,m_3(N_2,R_2,N_1))\nonumber\\
%&=&\omega( R_1,YXm_3(N_2,R_2,N_1))\nonumber\\
%&=&\omega(R_1,YM_3(N_2,R_2,N_1))\,.
\end{eqnarray}
Similarly, for two adjacent Ramond inputs, 
\begin{eqnarray}\label{cyc11}
\omega(N_1,M_3(R_1,R_2,N_2))&=&\omega(N_1,m_2(R_1,\mu_2(R_2,N_2)))-\omega(N_1,\mu_2(m_2(R_1,R_2),N_2))\\
&=&-\omega_L(N_1, m_2(\xi_0R_1,\mu_2(N_2,R_2)))-\omega_L(N_1,\mu_2(m_2(\xi_0 R_1,R_2),N_2))\,.\nonumber
\end{eqnarray}
Now, for the first term we use cyclicity of $m_2$ while for the second we use cyclicity of $\mu_2$ for two R-inputs which gives 
\begin{eqnarray}\label{3cyc11}
\omega(N_1,M_3(R_1,R_2,N_2))&=&\omega_L(\xi_0 R_1, m_2(\mu_2(R_2,N_2),N_1))+\omega_L(m_2(\xi_0R_1,R_2), \mu_2(N_2,N_1))\nonumber\\
&=&\omega_L(R_1, \xi_0 m_2(\mu_2(R_2,N_2),N_1))+\omega_L(R_1,\xi_0m_2(R_2, \mu_2(N_2,N_1)))\nonumber\\
&=&\omega(R_1,m_3(R_2,N_2,N_1))\,.
\end{eqnarray}
Thus,  $m_3$ is cyclic with respect to the symplectic form $\omega(\cdot,\cdot)$. 
%Along the same lines one shows that 
%\begin{eqnarray}
%\frac{1}{2}[M_2,M_2](N,R,R)&=&[Q,[\mu_2,M_2]](N,R,R)=:-[Q,M_3]](N,R,R)\nonumber\\
%\frac{1}{2}[M_2,M_2](N,N,R)&=&-\frac{1}{2}[Q,[M_2,\mu_2]](N,N,R)=:-[Q,M_3]](N,N,R)
%\end{eqnarray}
In order to prove cyclicity to arbitrary order we first recall the recursion relations defining the higher order products \cite{Erler:2015lya}. For zero or one Ramond input we have 
\begin{eqnarray}\label{1r}
M^{<0/1>}_{n+2}=\frac{1}{n+1}\sum\limits_{k=0}^n[M_{k+1},\mu_{n-k+2}]P_{n+2}^{<0/1>}\,,\qquad M_1=Q
\end{eqnarray}
and for two Ramond inputs%\footnote{Due to the choice of the homotopy (\ref{mun}) the recursion relation for the products $m'_n$ and the 'bare products' $m_n$ in \cite{Erler:2015lya} are unified in the present formalism.}
\begin{eqnarray}\label{2r}
M^{<2>}_{n+3}&=&m_{n+3} P_{n+3}^{<2>}=\frac{1}{n+1}\sum\limits_{k=0}^n[m_{k+2},\mu_{n-k+2}]P_{n+3}^{<2>}
\end{eqnarray}
where 
\begin{eqnarray}\label{2r2}
m_{n+3} =\frac{1}{n+1}\sum\limits_{k=0}^n[m_{k+2},\mu_{n-k+2}]
\end{eqnarray}
with $m_2=*$. Finally, the gauge products $\mu_{n}$ are given by 
\begin{eqnarray}\label{mun}
\mu_{n+2}=\frac{1}{n+3}\{{} \xi,m_{n+2}\}P_{n+2}^{<0>}+\xi m_{n+2} P_{n+2}^{<1>}\,.
\end{eqnarray}
It is not hard to see (by induction) that vanishing of $M_3(R,R,R)$ implies that vanishing of $M_{n+3}(\cdots,R,\cdots R,\cdots, R,\cdots)$ for all $n$. Indeed, upon inspection of (\ref{2r2}), subject to the homotopy (\ref{mun}), it is apparent that such a term would have to be of the form $\xi \sum\limits_{k=0}^nm_{n-k+2}m_{k+2}$ which vanishes due to the $A_\infty$ condition $[m,m]=0$. 
Furthermore, it holds that 
\begin{equation}\label{u1}
M_n^{<1>}=X\left(m_n^{<1>}\mu_2+m_{n-1}^{<1>}\mu_3+\cdots\right)=Xm_nP_n^{<1>}\,.
\end{equation}
To show this identity we proceed by induction. We have from (\ref{1r})
\begin{eqnarray}\label{ind}
nM_{n+1}^{<1>}&=&[M_{n}^{<1>},\mu_2^{<1>}]+[M_{n-1}^{<1>},\mu_3^{<1>}]+\cdots+[Q,\mu_{n+1}^{<1>}]\nonumber\\
&+&M_{n}^{<1>}\mu_2^{<0>}+M_{n-1}^{<1>}\mu_3^{<0>}+\cdots\nonumber\\
&-&\mu_2^{<1>}M_{n}^{<0>}+\mu_3^{<1>}M_{n-1}^{<0>}+\cdots\,.
\end{eqnarray}
Now, we use $[Q,\mu_p^{<1>}]=Xm_p^{<1>}-\xi[Q,m_p^{<1>}]$ together with the identity, $[m,M]=0$, that is,
\begin{eqnarray}\label{ind2}
[Q,\mu_{n+1}^{<1>}]&=&Xm_{n+1}^{<1>}+\xi\left([m_n^{<1>},M_2^{<1>}]+[m_{n-1}^{<1>},M_3^{<1>}]+\cdots\right.\nonumber\\
&+&\left.M_2^{<1>}m_n^{<0>}+M_3^{<1>}m_{n-1}^{<0>}+\cdots\right.\nonumber\\
&+&\left.m_n^{<1>}M_2^{<0>}+m_{n-1}^{<1>}M_3^{<0>}+\cdots\right)\,.
\end{eqnarray}
Upon substitution of (\ref{ind2}) into (\ref{ind}) and using (\ref{mun}) as well as $[m,m]=0$ the result follows. 

Thanks to (\ref{2r}) and (\ref{u1}) the problem of proving cyclicity of $M_n$ is again reduced to show cyclicity of $m_n$. To prove cyclicity of $m_{n+3}$, $n\geq 1$, one then proceeds exactly as in (\ref{cyc1})-(\ref{3cyc11}) expressing $m_{n+3}$ in terms of $[m_{k+2},\mu_{n-k+2}]$ and then using cyclicity of $m_q$, $q\leq n+2$ as well as cyclicity of $\mu_p$, $p\leq n+2$ for $p$ NS-inputs.

Let us now explain how these vertices lead to a gauge-invariant action for the open superstring in the small Hilbert space. Following \cite{Sen:2015uaa} we write 
%where the restriction on the Ramond fields is substituted by the introduction of an auxiliary field $\tilde \psi$ in the Ramond sector. In fact, our construction applies straight forwardly to this formulation as well. The only difference is that now,  $X=\tilde X $ is given by the expression originally used in \cite{Erler:2013xta}. %, that is,
%%\begin{equation}
%%X=\oint\frac{dz}{2\pi i} f(z) X(z)
%%\end{equation} 
%%and analogously for $\xi$. 
%The final open string action is then given by 
\begin{eqnarray}\label{so}
S&=&\frac{1}{2}\omega(\phi, Q\phi) -\frac{1}{2}\omega(\tilde\psi, XQ\tilde\psi) +\omega(\tilde\psi, Q\psi)\\ 
&&+\frac{1}{3}\omega(\Psi,{\cal{M}}_2(\Psi,\Psi))  +\frac{1}{4}\omega(\Psi,{\cal{M}}_3(\Psi,\Psi,\Psi))+\cdots \nonumber
\end{eqnarray}
where, $\Psi=\phi+\psi$ and $\tilde \psi$ is an auxiliary Ramond string field with picture $(-\frac{3}{2})$. The higher string products ${\cal{M}}_n$ are given by 
\begin{equation}
{\cal{M}}_n=M_nP^{<0>}+m_n(P^{<1>}+P^{<2>})
\end{equation}
which differs from (\ref{M2cyc}) by the ubiquitous factor $X$. To prove gauge invariance we use that $\mathcal{M}_n$ is cyclic w.r.t.~$\omega$. The standard proof of gauge-invariance has to be modified as $\mathcal{M}$ is not an $A_\infty$-algebra. However, $M$ is an $A_\infty$-algebra and differs from $\mathcal{M}$ in that it contains an additional $X$-insertion on Ramond outputs and contains no BRST operator $Q$. There are three different types of gauge-transformations with odd parameters $\Lambda$, $\lambda$ and $\tilde{\lambda}$ having picture $-1$, $-\frac{1}{2}$ and $-\frac{3}{2}$.

Using antisymmetry of $\omega$ and cyclicity of $\mathcal{M}_n$ one arrives at the identities ($n, k \geq 2$),
\begin{align}
 	\omega( \Lambda, M_n \circ M_k ) &= \omega( \Lambda, \mathcal{M}_n \circ M_k ) = \omega( \mathcal{M}_n {\boldsymbol \Lambda}, P_1^{<0>} \mathcal{M}_k + X P_1^{<1>} \mathcal{M}_k) = \omega( M_n {\boldsymbol \Lambda}, \mathcal{M}_k ), \label{eq:gb} \\
 	\omega( \Lambda, Q M_k ) &= \omega( Q \Lambda, M_k ) = \omega( Q {\boldsymbol \Lambda}, \mathcal{M}_k ), \\
 	\omega( \Lambda, M_n \circ Q ) &= \omega( \mathcal{M}_n {\boldsymbol \Lambda}, Q ). \label{eq:ge}
\end{align}
where ${\boldsymbol\Lambda}$ denotes the coderivation built from $\Lambda$ as its $0$-string map and we suppressed the string field $\Psi$. Explicitly, $(\ref{eq:ge})$ reads as
\begin{align*}
\omega( \Lambda, M_n( Q \Psi, \ldots, \Psi ) + M_n( \Psi, Q\Psi, \ldots, \Psi) + \cdots ) &=  \omega(\mathcal{M}_n( \Lambda, \Psi, \ldots, \Psi) + \mathcal{M}_n(\Psi, \Lambda,\ldots, \Psi) + \ldots, Q \Psi).
\end{align*}
Define the transformation $\delta \phi$, $\delta \psi$,  $\delta \tilde{\psi}$ as
\begin{align}
	\delta \phi + \delta{\tilde \psi} &= Q\Lambda + \sum_{n \geq 2} \mathcal{M}_n {\boldsymbol \Lambda} (e^{\Psi}), \label{eq:gbosb} \\
	\delta \psi &= X \delta \tilde{\psi}. \label{eq:gbose}
\end{align}
Summing over (\ref{eq:gb})-(\ref{eq:ge}) we obtain zero on the left-hand side due to the $A_\infty$ relations, while on the right-hand side we find,
\begin{align}
	0 &= \omega(\delta \phi, Q\phi ) + \omega(\delta \tilde{\psi}, Q \psi) + \sum_{k \geq 2} \omega( (\delta \phi + \delta \psi), \mathcal{M}_k(\Psi,\Psi,\ldots,\Psi)) \notag \\
	&= \delta \left( \frac{1}{2} \omega( \phi, Q \phi) + \omega(\tilde{\psi}, Q\psi) + \sum_{k \geq 2} \frac{1}{k+1} \omega(\Psi, \mathcal{M}_k(\Psi,\Psi,\ldots,\Psi)) \right) - \omega( \tilde{\psi}, Q \delta \psi ) \notag \\
	&= \delta S,
\end{align}
where we used $\omega(\tilde{\psi}, Q \delta \psi) = \delta \left(  \frac{1}{2} \omega( \tilde{\psi}, QX \tilde{\psi} ) \right)$ in the last step. Consequently, the transformations (\ref{eq:gbosb}) and (\ref{eq:gbose}) are a bosonic gauge symmetry of the action. By replacing $\Lambda$ with $\tilde{\lambda}$ in (\ref{eq:gb}) - (\ref{eq:ge}) one verifies that the following transformation is a fermionic gauge symmetry,
\begin{align}
	\delta \phi + \delta{\tilde \psi} &= Q\tilde{\lambda} + \sum_{n \geq 2} \mathcal{M}_n \mathbf{X} \tilde{\boldsymbol \lambda} (e^{\Psi}), \label{eq:gfer2b} \\
	\delta \psi &= X \delta \tilde{\psi}, \label{eq:gfer2e}
\end{align}
where $\mathbf{X} \tilde{\boldsymbol \lambda}$ denotes the coderivation with $0$-string product $X\lambda$.

In order to derive the gauge transformations corresponding to the parameter $\lambda$, let us recall that $M_n$ and $m_n (P^{<0>} + P^{<1>})$ give two commuting $A_\infty$ structures \cite{Erler:2015lya}. Together with cyclicity of $m_n (P^{<0>} + P^{<1>})$ w.r.t.~$\omega$ one can then deduce that the following transformations are a gauge symmetry of $S$, by imitating the previous derivation,
\begin{align}
	\delta \phi + \delta{\tilde \psi} &= \sum_{n \geq 2} \mathcal{M}_n {\boldsymbol \lambda} (e^{\Psi}), \label{eq:gferb} \\
	\delta \psi &= Q \lambda + X \delta \tilde{\psi}. \label{eq:gfere}
\end{align}
Notice that all gauge transformations preserve the constraint $\psi = X \tilde{\psi}$ up to states of the form $Q \lambda$ with $\lambda$ not expressible in the form $\lambda = X \rho$ for some picture $-\frac{3}{2}$ state $\rho$.

Let us now comment on the applicability of our formalism to writing the proposal for the superstring action \cite{Kunitomo:2015usa} in the small Hilbert space. Assuming the constraint (\ref{ryeh1}), we can rewrite (\ref{so}) without the need for the auxiliary field $\tilde\psi$ as 
\begin{equation}
S=\frac{1}{2}\omega(\phi, Q\phi) +\frac{1}{2}\omega(\psi, YQ\psi) +\frac{1}{3}\omega(\Psi,{\cal{M}}_2(\Psi,\Psi))  +\frac{1}{4}\omega(\Psi,{\cal{M}}_3(\Psi,\Psi,\Psi))+\cdots 
\end{equation}
where $Y=c_0\delta'(\gamma_0)$ is the inverse picture changing operator in the restricted Hilbert space. The gauge transformation of this action agrees with that of (\ref{so}) up to the contribution coming from the kinetic term that is 
\begin{equation}
\delta S\propto \omega((X-X_0)(m_2(\Psi,\Lambda)+m_2(\Lambda,\Psi)+m_3(\Psi,\Lambda,\Psi+\cdots)),YQ\psi)
\end{equation}
Formally this term can be removed by replacing $X$ by $X_0$ (as well as $\xi$ by 
$\Theta(\beta_0)$) in the definition of the higher string products $M_n$ and the gauge products $\mu_n$ when applied to states containing one or two Ramond states, e.g. instead of (\ref{M2cyc}) we take 
\begin{eqnarray}
M_2=\frac{1}{3}\{{} X,m_2\}P_2^{<0>}+ X_0m_2 P_2^{<1>}+m_2 P_2^{<2>}
\end{eqnarray}
and instead of (\ref{mu2}) we take 
\begin{equation}
\mu_2=\Theta(\beta_0) m_2 P_2^{<1>}+\frac{1}{3}\{{} \xi,m_2\}P_2^{<0>}\,.
\end{equation}
However, for this choice of homotopy to be well defined, one needs that the $m_n$s are compatible with the particular realisation of the picture $(-\frac{1}{2})$ states in terms of the zero modes $\beta_0$ and $\gamma_0$ described in section 1.

 \bigskip
\noindent{\bf Acknowledgements:}\\
\smallskip
We would like to thank Ted Erler and Barton Zwiebach for helpful discussions. I.S. would like to thank the Center for the Fundamental Laws of Nature at Harvard University for hospitality during the initial stages of this work. This work  was supported by the DFG Transregional Collaborative Research Centre TRR 33 and the DFG cluster of  excellence "Origin and Structure of the Universe".

\end{document}